\documentclass[pra,aps,nofootinbib,notitlepage,superscriptaddress,twocolumn]{revtex4-1}
\usepackage{amsthm}
\usepackage{amsmath,bm}
\usepackage{amssymb}
\usepackage{amsfonts}
\usepackage{graphicx}
\usepackage{fancyhdr}
\usepackage{txfonts}
\usepackage[vcentermath]{youngtab}
\usepackage[colorlinks=true,linkcolor=blue,citecolor=blue,urlcolor=blue]{hyperref}

\begin{document}

\title{$PT$-symmetric non-Hermitian Hamiltonian and invariant operator in
periodically driven $SU(1,1)$ system}
\author{Yan Gu}
\affiliation{Institute of Theoretical Physics and Department of Physics, State Key
Laboratory of Quantum Optics and Quantum Optics Devices, Shanxi University,
Taiyuan, Shanxi 030006, China.}
\author{Xue-Min Bai}
\affiliation{School of Physics, Jinzhong University, Jinzhong 030619,
Shanxi, China.}
\author{Xiao-Lei Hao}
\author{J. -Q. Liang}
\email{jqliang@sxu.edu.cn}
\affiliation{Institute of Theoretical Physics and Department of Physics, State Key
Laboratory of Quantum Optics and Quantum Optics Devices, Shanxi University,
Taiyuan, Shanxi 030006, China.}
\begin{abstract}
We study in this paper the time evolution of $PT$-symmetric non-Hermitian
Hamiltonian consisting of periodically driven $SU(1,1)$ generators. A
non-Hermitian invariant operator is adopted to solve the Schr\"{o}dinger
equation, since the time-dependent Hamiltonian is no longer a conserved
quantity. We propose a scheme to construct the non-Hermitian invariant with
a $PT$-symmetric but non-unitary transformation operator. The eigenstates of
invariant and its complex conjugate form a bi-orthogonal basis to formulate
the exact solution. We obtain the non-adiabatic Berry phase, which reduces
to the adiabatic one in the slow time-variation limit. A non-unitary time-
evolution operator is found analytically. As an consequence of the
non-unitarity the ket ($|\psi (t)\rangle $) and bra ($\langle \psi (t)|$)
states are not normalized each other. While the inner product of two states
can be evaluated with the help of a metric operator. It is shown explicitly
that the model can be realized by a periodically driven oscillator.
\end{abstract}

\keywords{$PT$-symmetry, Berry phase, non-Hermitian invariant.}
\pacs{11.30.Er; 03.65.Fd; 03.65.Vf}
\date{[]}
\maketitle

\volumeyear{year} \volumenumber{number} \issuenumber{number} \eid{identifier}

\received[Received text]{date}

\revised[Revised text]{date}

\accepted[Accepted text]{date}

\published[Published text]{date}

\startpage{1} \endpage{2} \preprint{ }


%\accepted{(Day Month Year)}
%\comby{(xxxxxxxxxx)}
\thispagestyle{fancy}

\section{Introduction}

In 1998, Bender and Boettcher showed that the Hermiticity of Hamiltonian is
sufficient but not necessary condition to have real spectrum \cite%
{Bender1998}. The non-Hermitian Hamiltonians can still possess real and
positive eigenvalues \cite{Bender1998}, provided that the parity-time
symmetry ($PT$-symmetry) is maintained instead of Hermiticity. Soon after
the originally proposed Hamiltonians \cite{Bender1998,Bender2002,Bender2007}
have been extended to different kinds of $PT$-symmetric non-Hermitian
systems. The predicted properties of $PT$ -symmetric Hamiltonians\cite%
{Bender2007,Most2010} have been observed at the classical level in a wide
variety of laboratory experiments involving superconductivity\cite%
{Rubi2007,Chtc2012}, optics\cite{Guo2009,Ruter2010,Lin2011,Feng2011},
microwave cavities\cite{Bittner2012}, atomic diffusion\cite{Zhao2010}, and
nuclear magnetic resonance\cite{Zheng2013}.

The study of $PT$ symmetric oscillator has attracted much attention in
recent years. In 2005, Cem Yuce studied the exactly solvable generalized $PT$
symmetric harmonic oscillator problem\cite{Yuce2005}. Joseph Schindler $et$ $%
al$. presented in 2011 a simple experimental set-up, which displays all the
novel phenomena of $PT$ symmetry for a coupled oscillator pair\cite%
{Schindler2011}. Subsequently Carl M. Bender $et$ $al$. observed\ $PT$ phase
transition in a $PT$-symmetric two-oscillator model\cite{Benderajp}, and
studied twofold transitions as well\cite{Bender2013}. A systematic analysis
is provided by Jes\'{u}s Cuevas $et$ $al$ for a prototypical nonlinear
oscillator with $PT$-symmetry\cite{Cuevas2013}. The phase transition arises
when the number of coupled oscillator-pairs increases from $1$ to $N$, which
is sufficient large\cite{Bender2014}. A partial $PT$ symmetry is examined by
Alireza Beygi $et$ $al$. for a chain of $N$ coupled harmonic-oscillators\cite%
{Beygi2015}. Exact analytical solutions are found by Andreas Fring and
Thomas Frith for a two-dimensional time-dependent non-Hermitian describing
coupled two harmonic oscillators, which possess infinite dimensional Hilbert
space in the broken $PT$-symmetry regime\cite{Fring2018}. And then they
provided a time-dependent Dyson map and metric recently\cite{Fring2020}. The
dynamics of the average displacement of a mechanical oscillator is also
investigated in different regimes for the $PT$-symmetric-like optomechanical
system\cite{Xu2021}.

Quantum harmonic oscillator is exactly solvable and effectively applied to
many other systems\cite{Dirac1958}, for example, a cavity mode in cavity
quantum electrodynamics \cite{Raus2000} or a mode of an LC radio-frequency
resonator in circuit quantum electrodynamics\cite{Wallraff2004}.
Entanglement dynamics is researched in short- and long-range harmonic
oscillators\cite{Nezh2014}.

Since the time dependent Hamiltonian is no longer a conserved quantity it is
necessary to construct a $PT$ symmetric non-Hermitian invariant operator to
solve the time dependent system. The Hermitian invariants were proposed by
Lewis and Riesenfeld to investigate the dynamics and quantization of
time-dependent systems long ago \cite{Lewis1968,Lewis1969}. Explicit
Hermitian invariant operators are constructed by Y. Z. Lai $et$ $al$ for the
time-dependent quantum systems consisting of $SU\left( 1,1\right) $ and $%
SU\left( 2\right) $ generators\cite{Laipra}. B. Khantoul $et$ $al$. propose
a scheme to deal with certain time-dependent non-Hermitian Hamiltonian,
which involves the use of invariant operators, which are pseudo-Hermitian
with respect to the time-dependent metric operator\cite{Khantoul2017}.
Topological invariants are also discussed in non-Hermitian systems\cite%
{Liang2013,Yao2018,Ghatak2019,Song2019}.

For the non-Hermitian Hamiltonian the invariant operators have to be $PT$%
-symmetric and non-Hermitian. Subsequently $PT$-symmetric transformations
are required to construct the invariant operators instead of the usual
unitary transformation in the system described by the Hermitian Hamiltonian
\cite{Laipra}. Therefore, the main aim of this paper is to invite an
alternative method with $PT$-symmetric non-Hermitian invariant operators for
the time-evolution solution of the non-Hermitian Hamiltonian.

If the time dependence of the Hamiltonian depends on a set of parameters, a
cyclic evolution of the Hamiltonian in the parameter space leads to an
additional phase that has geometric significance and is known as Berry's
phase\cite{Berry1984}. This phase shift, which reveals a gauge structure in
quantum mechanics, has attracted both theoretical and experimental interests%
\cite{Zhang2005,Watanabe2018,Enomoto2019}. We obtain the nonadiabatic Barry
phase, which reduces to the adiabatic one in the slow time-varying limit for
the non-Hermitian Hamiltonian.

The paper is organized as follows: in Sec. II we put forward a $PT$%
-symmetric non-Hermitian Hamiltonian consisting of periodically driven $%
SU(1,1)$ generators. The $PT$-symmetric non-Hermitian invariant operator is
constructed by means of $PT$-symmetric transformation in order to obtain the
time evolution of quantum states. We explain bi-orthogonal basis approach
and metric operator in Sec. III. Exact solutions of the Schr\"{o}dinger
equations are found along with the nonadiabatic Berry phase, which reduces
the adiabatic one in the slow varying limit. The time-evolution operator is
presented in Sec. IV. We end with a conclusion and discussion in the last
section.

\section{$PT$-symmetric non-Hermitian Hamiltonian and invariant operator}

In classical mechanics $PT$ transformation is simply the reflections of
space and time coordinates. While the $PT$ transformation operator in
quantum mechanics can be defined as
\begin{equation*}
\hat{\Theta}=\hat{U}K,\hat{\Theta}^{-1}=K\hat{U}^{\dag }
\end{equation*}%
where $\hat{U}$ is the usual unitary operator for the space-time reflection
and $K$ denotes the operation of complex conjugate. The position and
momentum operators become $\hat{\Theta}\hat{x}\hat{\Theta}^{-1}=-\hat{x}$, $%
\hat{\Theta}\hat{p}\hat{\Theta}^{-1}=\hat{p}$ under $PT$ transformation, and
$\hat{\Theta}\hat{a}\hat{\Theta}^{-1}=-\hat{a},\ \hat{\Theta}\hat{a}^{\dag }%
\hat{\Theta}^{-1}=-\hat{a}^{\dag }$ for the boson operators in harmonic
oscillator$.$The $PT$ transformation operator is antilinear and antiunitary.
The $SU(1,1)$ generators satisfy the commutation relation%
\begin{equation}
\left[ \hat{S}_{+},\hat{S}_{-}\right] =-2\hat{S}_{z},\text{ }\left[ \hat{S}%
_{z},\hat{S}_{\pm }\right] =\pm \hat{S}_{\pm }.  \label{1}
\end{equation}%
And the $SU(1,1)$ Lie algebra has a realization in terms of boson creation
and annihilation operators $\hat{a}^{\dag }$\ and $\hat{a}$ such that\cite%
{Laipra,Liang2020}%
\begin{equation}
\hat{S}_{z}=\frac{1}{2}\left( \hat{a}^{\dag }\hat{a}+\frac{1}{2}\right) ,%
\hat{S}_{+}=\frac{1}{2}\left( \hat{a}^{\dag }\right) ^{2},\hat{S}_{-}=\frac{1%
}{2}\left( \hat{a}\right) ^{2},
\end{equation}%
Under the $PT$ transformation the $SU(1,1)$ generators transform as%
\begin{equation}
\hat{\Theta}\hat{S}_{z}\hat{\Theta}^{-1}=\hat{S}_{z},\ \hat{\Theta}\hat{S}%
_{+}\hat{\Theta}^{-1}=\hat{S}_{+},\ \hat{\Theta}\hat{S}_{-}\hat{\Theta}^{-1}=%
\hat{S}_{-}.
\end{equation}%
The commutation relation Eq.(\ref{1}) is $PT$ invariant.

We consider the $PT$-symmetric non-Hermitian Hamiltonian written as%
\begin{equation}
\hat{H}\left( t\right) =\Omega \hat{S}_{z}+G\left( \hat{S}_{+}e^{i\phi
\left( t\right) }-\hat{S}_{-}e^{-i\phi \left( t\right) }\right)  \label{N1}
\end{equation}%
in which
\begin{equation*}
\phi \left( t\right) =\omega t
\end{equation*}%
with $\omega $ being the driving frequency and $G$ denotes a coupling
parameter. The Hamiltonian is obviously non-Hermitian
\begin{equation*}
\hat{H}\left( t\right) \neq \hat{H}^{\dag }\left( t\right) =\Omega \hat{S}%
_{z}-G\left( \hat{S}_{+}e^{i\phi \left( t\right) }-\hat{S}_{-}e^{-i\phi
\left( t\right) }\right) ,
\end{equation*}%
since $\hat{S}_{+}=\left( \hat{S}_{-}\right) ^{\dag }$. As a matter of fact
the Hamiltonian $\hat{H}\left( t\right) $ describes a periodically driving
harmonic-oscillator
\begin{eqnarray*}
\hat{H}\left( t\right) &=&\left[ \frac{\Omega }{4}+i\frac{G}{2}\sin \phi
\left( t\right) \right] \hat{x}^{2}+\left[ \frac{\Omega }{4}-i\frac{G}{2}%
\sin \phi \left( t\right) \right] \hat{p}^{2} \\
&&-i\frac{G}{2}\cos \phi \left( t\right) \left( \hat{x}\hat{p}+\hat{p}\hat{x}%
\right) ,
\end{eqnarray*}%
in the coordinate and momentum representation of boson operators $\hat{a}%
=\left( \hat{x}+i\hat{p}\right) /\sqrt{2}$ and $\hat{a}^{\dag }=\left( \hat{x%
}-i\hat{p}\right) /\sqrt{2}$. The Schr\"{o}dinger equation is covariant
under the $PT$ transformation
\begin{equation*}
i\frac{d}{dt}\left\vert \psi ^{\prime }\left( t\right) \right\rangle =\hat{H}%
\left( t\right) \left\vert \psi ^{\prime }\left( t\right) \right\rangle ,
\end{equation*}%
(natural unit $\hbar =1$) with $\left\vert \psi ^{\prime }\left( t\right)
\right\rangle =\hat{\Theta}\left\vert \psi \left( t\right) \right\rangle $.
We can solve the time-dependent $SU(1,1)$ system with non-Hermitian
Hamiltonian in the formalism of Schr\"{o}dinger equation.

Since the Hamiltonian is no long a conserved quantity an invariant operator
is required to solve the time dependent system\cite%
{Lewis1968,Lewis1969,Laipra}. For the $PT$-symmetric Hamiltonian the
invariant operator $\hat{I}\left( t\right) $, which satisfies the condition
\begin{equation}
i\frac{d\hat{I}\left( t\right) }{dt}=i\frac{\partial }{\partial t}\hat{I}%
\left( t\right) +\left[ \hat{I}\left( t\right) ,\hat{H}\left( t\right) %
\right] =0,  \label{invar}
\end{equation}%
should also be invariant under the $PT$ transformation. To this end we can
construct $\hat{I}\left( t\right) $ from the $PT$-symmetric $SU(1,1)$%
-generator $\hat{S}_{z}$\ with a $PT$-symmetric transformation such that%
\begin{equation}
\hat{I}\left( t\right) =\hat{R}\left( t\right) \hat{S}_{z}\hat{R}^{-1}\left(
t\right) .  \label{I}
\end{equation}%
The transformation operator considered as%
\begin{eqnarray}
\hat{R}\left( t\right) &=&e^{\frac{\eta }{2}\left( \hat{S}_{+}e^{i\phi
\left( t\right) }+\hat{S}_{-}e^{-i\phi \left( t\right) }\right) },  \label{R}
\\
\hat{R}^{-1}\left( t\right) &=&e^{-\frac{\eta }{2}\left( \hat{S}_{+}e^{i\phi
\left( t\right) }+\hat{S}_{-}e^{-i\phi \left( t\right) }\right) },  \notag
\end{eqnarray}%
(with a real parameter $\eta $ to be determined) is $PT$ symmetric, since%
\begin{equation*}
\hat{\Theta}\hat{R}\left( t\right) \hat{\Theta}^{-1}=\hat{R}\left( t\right) .
\end{equation*}%
However the operator $\hat{R}\left( t\right) $ is non-unitary with $\hat{R}%
^{\dag }\neq \hat{R}^{-1}$ different from the ordinary quantum mechanics.
Occasionally it is Hermitian ($\hat{R}^{\dag }(t)$=$\hat{R}\left( t\right) $%
) in this particular model but is unnecessary in general. The invariant
operator $\hat{I}\left( t\right) $ is $PT$-symmetric, however non-Hermitian
\begin{equation*}
\hat{I}\left( t\right) \neq \hat{I}^{\dag }\left( t\right) =\hat{R}%
^{-1}\left( t\right) \hat{S}_{z}\hat{R}\left( t\right) .
\end{equation*}%
With the $PT$-symmetric transformation operator Eq.(\ref{R}) we obtain the
following relations\cite{Lai1996}%
\begin{eqnarray}
\hat{R}^{-1}\left( t\right) \hat{S}_{+}\hat{R}\left( t\right) &=&\hat{S}%
_{+}\cos ^{2}\left( \frac{\eta }{2}\right) -\hat{S}_{z}e^{-i\phi \left(
t\right) }\sin \left( \eta \right)  \notag \\
&&+\hat{S}_{-}e^{-2i\phi \left( t\right) }\sin ^{2}\left( \frac{\eta }{2}%
\right) ,  \notag \\
\hat{R}^{-1}\left( t\right) \hat{S}_{-}\hat{R}\left( t\right) &=&\hat{S}%
_{-}\cos ^{2}\left( \frac{\eta }{2}\right) +\hat{S}_{z}e^{i\phi \left(
t\right) }\sin \left( \eta \right)  \notag \\
&&+\hat{S}_{+}e^{2i\phi \left( t\right) }\sin ^{2}\left( \frac{\eta }{2}%
\right) ,  \notag \\
\hat{R}^{-1}\left( t\right) \hat{S}_{z}\hat{R}\left( t\right) &=&\hat{S}%
_{z}\cos \left( \eta \right)  \notag \\
&&+\frac{1}{2}\sin \left( \eta \right) \left( \hat{S}_{+}e^{i\phi \left(
t\right) }-\hat{S}_{-}e^{-i\phi \left( t\right) }\right)  \label{S1}
\end{eqnarray}%
and%
\begin{eqnarray}
i\hat{R}^{-1}\left( t\right) \frac{\partial }{\partial t}\hat{R}\left(
t\right) &=&2\frac{d\phi }{dt}\hat{S}_{z}\sin ^{2}\left( \frac{\eta }{2}%
\right)  \notag \\
&&-\frac{d\phi }{2dt}\sin \left( \eta \right) \left( \hat{S}_{+}e^{i\phi
\left( t\right) }-\hat{S}_{-}e^{-i\phi \left( t\right) }\right)  \label{S2}
\end{eqnarray}

By straightforward algebra the invariant operator $\hat{I}\left( t\right) $
is found as\cite{Lai1996}%
\begin{equation}
\hat{I}\left( t\right) =\cos \left( \eta \right) \hat{S}_{z}-\frac{1}{2}\sin
\left( \eta \right) \left( \hat{S}_{+}e^{i\phi \left( t\right) }-\hat{S}%
_{-}e^{-i\phi \left( t\right) }\right) ,
\end{equation}%
under the auxiliary condition%
\begin{equation}
G\cos \left( \eta \right) =-\frac{1}{2}(\dot{\phi}+\Omega )\sin \left( \eta
\right) ,  \label{AS}
\end{equation}%
derived from Eq.(\ref{invar}), Eq.(\ref{I}) and Eq.(\ref{R}).

\section{Exact solution and non-adiabatic Berry phase}

Since the invariant operator $\hat{I}(t)$ is non-Hermitian the eigenstates
of it along are not an orthonormal basis. The bi-orthogonal basis\cite%
{Sun1993,Leung1998,Shi2009} are requested respectively for the $PT$%
-symmetric invariant-operator $\hat{I}(t)$ and its complex conjugate $\hat{I}%
^{\dag }\left( t\right) $ to obtain the exact solution of the Schr\"{o}%
dinger equation.

\subsection{Bi-orthogonal basis and metric operator}

The eigenstates of $\hat{S}_{z}$, $\hat{S}_{z}\left\vert n\right\rangle
=k_{n}\left\vert n\right\rangle ,$ are Fock states $\left\vert
n\right\rangle $ with eigenvalues
\begin{equation*}
k_{n}=\frac{1}{2}\left( n+\frac{1}{2}\right) ,
\end{equation*}%
which are nothing but the eigenvalues of harmonic oscillator. The
eigenstates of $\hat{I}(t)$ are obviously given by%
\begin{equation*}
\hat{I}(t)\left\vert n(t)\right\rangle _{r}=k_{n}\left\vert
n(t)\right\rangle _{r},\text{\ }\left\vert n(t)\right\rangle _{r}=\hat{R}%
\left( t\right) \left\vert n\right\rangle ,
\end{equation*}%
with the same eigenvalues $k_{n}$, which are conserved quantities. The
subscript "$r$" denotes the ket states. Since the $PT$-symmetric
transformation operator $\hat{R}\left( t\right) $ is non-unitary, the
states\ $\left\vert n\left( t\right) \right\rangle _{r}$ cannot be
normalized i. e. $_{r}\left\langle n\left( t\right) \right\vert n\left(
t\right) \rangle _{r}\neq 1$. The complex conjugate operator of invariant $%
\hat{I}(t)$ is seen to be%
\begin{equation*}
\hat{I}^{\dag }\left( t\right) =\left[ \hat{R}\left( t\right) \hat{S}_{z}%
\hat{R}^{-1}\left( t\right) \right] ^{\dag }=\hat{R}^{-1}\left( t\right)
\hat{S}_{z}\hat{R}\left( t\right) ,
\end{equation*}%
which possesses eigenstates denoted by%
\begin{equation*}
\hat{I}^{\dag }\left( t\right) \left\vert n(t)\right\rangle
_{l}=k_{n}\left\vert n(t)\right\rangle _{l},\text{\ }\left\vert
n(t)\right\rangle _{l}=\hat{R}^{-1}\left( t\right) \left\vert n\right\rangle
.
\end{equation*}%
with real eigenvalues too and the subscript "$l$" indicates the bra states
in the orthogonality condition. The two sets of the ket and bra basis \{$%
\left\vert n\left( t\right) \right\rangle _{r}$\} and \{$\left\vert n\left(
t\right) \right\rangle _{l}$\} form a bi-orthogonal basis\cite%
{Sun1993,Leung1998,Shi2009} with the orthogonality condition
\begin{equation}
_{l}\left\langle n(t)\right\vert m(t)\rangle _{r}=\delta _{nm}.  \label{o}
\end{equation}%
The completeness relation is%
\begin{equation*}
\sum_{n}\left\vert n(t)\right\rangle _{rl}\left\langle n(t)\right\vert
=\sum_{n}\left\vert n(t)\right\rangle _{lr}\left\langle n(t)\right\vert =1.
\end{equation*}%
According to Refs.\cite{Shi2009}, we can define new metric operator $\hat{%
\chi}$ relating the ket and bra states by%
\begin{equation*}
\left\vert n(t)\right\rangle _{l}=\hat{\chi}\left\vert n(t)\right\rangle
_{r},
\end{equation*}%
in which the metric operator for the present model is%
\begin{equation}
\hat{\chi}=\left( \hat{R}^{-1}\left( t\right) \right) ^{2}.  \label{m}
\end{equation}%
With the metric operator orthogonality condition Eq.(\ref{o}) becomes
\begin{equation*}
(n(t),m(t))\equiv \left\langle n(t)\right\vert \hat{\chi}\left\vert
m(t)\right\rangle =\delta _{nm}
\end{equation*}%
without using the two sets of basis.

\subsection{Solution and LR phase}

According to Lewis and Riesenfeld (LR) theory\cite{Lewis1968,Lewis1969}, the
general solution of the Schr\"{o}dinger equation is superposition of the
eigenstates of invariant operator $\hat{I}(t)$
\begin{equation}
\left\vert \psi \left( t\right) \right\rangle =\sum_{n}C_{n}e^{i\alpha
_{n}\left( t\right) }\left\vert n(t)\right\rangle _{r},  \label{g}
\end{equation}%
in which the time-independent coefficient $C_{n}$ can be determined by
initial condition. Under the $PT$-symmetric transformation
\begin{equation*}
\left\vert \psi ^{\prime }\left( t\right) \right\rangle =\hat{R}%
^{-1}\left\vert \psi \left( t\right) \right\rangle ,
\end{equation*}%
the original time-dependent Schr\"{o}dinger equation $i\frac{\partial }{%
\partial t}\left\vert \psi \left( t\right) \right\rangle =\hat{H}\left\vert
\psi \left( t\right) \right\rangle $ becomes
\begin{equation}
i\frac{\partial }{\partial t}\left\vert \psi ^{\prime }\left( t\right)
\right\rangle =\hat{H}^{\prime }\left\vert \psi ^{\prime }\left( t\right)
\right\rangle  \label{s}
\end{equation}%
with the new Hamiltonian given by%
\begin{equation*}
\hat{H}^{\prime }=\left( \hat{R}^{-1}\hat{H}\hat{R}-i\hat{R}^{-1}\frac{%
\partial }{\partial t}\hat{R}\right) .
\end{equation*}%
The solution of Schr\"{o}dinger equation Eq.(\ref{s}) is
\begin{equation}
\left\vert \psi ^{\prime }\left( t\right) \right\rangle
=\sum_{n}C_{n}e^{i\alpha _{n}\left( t\right) }\left\vert n\right\rangle ,
\label{l}
\end{equation}%
seen from Eq.(\ref{g}). Substituting the $\left\vert \psi ^{\prime }\left(
t\right) \right\rangle $ in Eq.(\ref{l}) into the Schr\"{o}dinger equation (%
\ref{s}) yields the LR phase
\begin{eqnarray}
\alpha _{n}\left( t\right) &=&-\int_{0}^{t}dt^{\prime }\left\langle
n\right\vert \hat{H}^{\prime }\left( t^{\prime }\right) \left\vert
n\right\rangle  \notag \\
&=&\int_{0}^{t}dt_{l}^{\prime }\left\langle n(t^{\prime })\right\vert \left[
i\frac{\partial }{\partial t^{\prime }}-\hat{H}\left( t^{\prime }\right) %
\right] \left\vert n(t^{\prime })\right\rangle _{r}.  \label{L}
\end{eqnarray}%
The periodically driven Hamiltonian can be solved exactly by means of the LR
method with the help of invariant operator. The first term in LR phase $%
\alpha _{n}\left( t\right) $ is usually regarded as the non-adiabatic Berry
phase. Using Eqs.(\ref{S1},\ref{S2}) we obtain the LR phase
\begin{equation}
\alpha _{n}\left( t\right) =-k_{n}\int_{0}^{t}\left[ \Omega -2\Gamma \left(
t^{\prime }\right) \right] dt^{\prime },  \label{b}
\end{equation}%
where $\Gamma \left( t\right) $ is given by%
\begin{equation}
\Gamma =\left( \dot{\phi}+\Omega \right) \sin ^{2}\left( \frac{\eta }{2}%
\right) +G\sin \left( \eta \right) .  \label{c}
\end{equation}

From the auxiliary equation Eq.(\ref{AS}), the parameter $\eta $ is
determined as%
\begin{equation}
\sin ^{2}\left( \frac{\eta }{2}\right) =\frac{1}{2}\mp \frac{\omega +\Omega
}{2\sqrt{(\omega +\Omega )^{2}+4G^{2}}}.  \label{n}
\end{equation}

\subsection{Berry phase}

According to definition, the first term in Eq.(\ref{L}) gives rise to the
Berry phase indicated by $\gamma _{n}$, which in one period of the driven
field $T=2\pi /\omega $ is evaluated from Eq.(\ref{S2}) as%
\begin{equation}
\gamma _{n}\left( T\right) =i\int_{0}^{T}{}_{l}\left\langle n\left( t\right)
\right\vert \frac{\partial }{\partial t}\left\vert n\left( t\right)
\right\rangle _{r}dt=2k_{n}\oint \sin ^{2}\left( \frac{\eta }{2}\right)
d\phi .  \label{Berry}
\end{equation}%
Substituting Eq.(\ref{n}) into Eq.(\ref{Berry}) we find the non-adiabatic
Berry phase
\begin{equation*}
\gamma _{n}\left( T\right) =\pi \left( n+\frac{1}{2}\right) \left( 1\mp
\frac{\omega +\Omega }{\sqrt{(\omega +\Omega )^{2}+4G^{2}}}\right) .
\end{equation*}%
In the adiabatic approximation that%
\begin{equation*}
\dot{\phi}=\omega =0
\end{equation*}%
the Berry phase becomes the well known form%
\begin{equation*}
\gamma _{n}\left( T\right) =\pi \left( n+\frac{1}{2}\right) \left( 1\mp
\frac{\Omega }{\sqrt{\Omega ^{2}+4G^{2}}}\right) .
\end{equation*}

\subsection{Average energies}

The invariant operators $\hat{I}(t)$ and its conjugate $\hat{I}^{\dag
}\left( t\right) $ possess real eigenvalues in the $PT$ -symmetric $SU(1,1)$
system. From Eqs.(\ref{L},\ref{b},\ref{c}), one can easily confirm that the
transformed Hamiltonian does not depend on time%
\begin{equation*}
\hat{R}^{-1}\left( t\right) \hat{H}\left( t\right) \hat{R}\left( t\right)
=\left( \Omega -2\Gamma _{0}\right) \hat{S}_{z}
\end{equation*}%
in which%
\begin{equation*}
\Gamma _{0}=\Omega \sin ^{2}\left( \frac{\eta }{2}\right) +G\sin \left( \eta
\right)
\end{equation*}%
in the adiabatic approximation. Average energies at the eigenstates of
invariant operators $\hat{I}(t)$ and its conjugate $\hat{I}^{\dag }\left(
t\right) $ are entirely real values too%
\begin{equation*}
_{l}\left\langle n(t)\right\vert \hat{H}\left( t\right) \left\vert
n(t)\right\rangle _{r}=\left\langle n\right\vert \left( \Omega -2\Gamma
_{0}\right) \hat{S}_{z}\left\vert n\right\rangle =\left( \Omega -2\Gamma
_{0}\right) k_{n}.
\end{equation*}%
The Hamiltonian $\hat{H}\left( t\right) $ what we considered is in an
unbroken $PT$-symmetric phase according to Ref.(\cite{Benderajp}).

\section{The time-evolution operator}

The time-evolution operator can be derived in terms of the eigenstates of
invariant operator. Substituting the LR phase Eq.(\ref{b}) into the general
solution of the Schr\"{o}dinger equation (\ref{g}) we have%
\begin{equation}
\left\vert \psi \left( t\right) \right\rangle =\hat{R}\left( t\right)
\sum_{n}C_{n}e^{-i\epsilon \left( t\right) k_{n}}\left\vert n\right\rangle =%
\hat{R}\left( t\right) e^{-i\epsilon \left( t\right) \hat{S}%
_{z}}\sum_{n}C_{n}\left\vert n\right\rangle ,  \label{g1}
\end{equation}%
where%
\begin{equation*}
\epsilon \left( t\right) =\int_{0}^{t}dt^{\prime }\left[ \Omega -2\Gamma
\left( t^{\prime }\right) \right] .
\end{equation*}%
Assume that the initial state at time $t=0$ is denoted by%
\begin{equation*}
\left\vert \psi \left( 0\right) \right\rangle =\hat{R}\left( 0\right)
\sum_{n}C_{n}\left\vert n\right\rangle
\end{equation*}%
The state at time $t$ is generated by the time evolution operator such that%
\begin{equation*}
\left\vert \psi \left( t\right) \right\rangle =\hat{U}\left( t,0\right)
\left\vert \psi \left( 0\right) \right\rangle ,
\end{equation*}%
where the time evolution operator is
\begin{equation}
\hat{U}\left( t,0\right) =\hat{R}\left( t\right) e^{-i\epsilon \left(
t\right) \hat{S}_{z}}\hat{R}^{-1}\left( 0\right) .  \label{u}
\end{equation}%
The time evolution operator is not unitary
\begin{equation*}
\hat{U}^{-1}\left( t,0\right) \neq \hat{U}^{\dag }\left( t,0\right) ,
\end{equation*}%
since the inverse operator is%
\begin{equation*}
\hat{U}^{-1}\left( t,0\right) =\hat{R}\left( 0\right) e^{i\epsilon \left(
t\right) \hat{S}_{z}}\hat{R}^{-1}\left( t\right)
\end{equation*}%
while the complex conjugate reads%
\begin{equation*}
\hat{U}^{\dag }\left( t,0\right) =\hat{R}^{-1}\left( 0\right) e^{i\epsilon
\left( t\right) \hat{S}_{z}}\hat{R}\left( t\right) .
\end{equation*}%
The bra state
\begin{equation*}
\left\langle \psi \left( t\right) \right\vert =\left\langle \psi \left(
0\right) \right\vert \hat{U}^{\dag }\left( t,0\right) ,
\end{equation*}%
is not normalized with the corresponding ket state%
\begin{equation*}
\left\langle \psi \left( t\right) \right\vert \psi \left( t\right) \rangle
\neq 1
\end{equation*}%
While it can be normalized with the help of metric operator\cite{Shi2009}%
\begin{equation*}
\left( \psi \left( t\right) ,\psi \left( t\right) \right) \equiv
\left\langle \psi \left( t\right) \right\vert \hat{\chi}\left( t\right)
\left\vert \psi \left( t\right) \right\rangle =1.
\end{equation*}%
In general the inner product of two states $\left\vert \psi \left( t\right)
\right\rangle $ and $\left\vert \varphi \left( t\right) \right\rangle $ is
evaluated as%
\begin{equation*}
\left( \varphi \left( t\right) ,\psi \left( t\right) \right) \equiv
\left\langle \varphi \left( t\right) \right\vert \hat{\chi}\left( t\right)
\left\vert \psi \left( t\right) \right\rangle .
\end{equation*}

\section{Conclusion}

The $PT$ symmetry is a more general property of dynamic system, since Both
Newton's and Maxwell's equations are invariant under the $PT$
transformation. If the non-Hermitian Hamiltonian is $PT$ symmetric, the Schr%
\"{o}dinger equation is also invariant. We demonstrate an analytical
formalism to solve the periodically driven non-Hermitian Hamiltonian
consisting of generators of $SU(1,1)$ Lie algebra. A $PT$-symmetric
non-Hermitian invariant operator $\hat{I}\left( t\right) $ is constructed in
terms of $PT$-symmetric but non-unitary transformation-operator $\hat{R}(t)$%
. As a consequence two sets of basis \{$\left\vert n\left( t\right)
\right\rangle _{r}$\}, \{$\left\vert n\left( t\right) \right\rangle _{l}$\}
respectively for invariant $\hat{I}(t)$ and its complex conjugate operator $%
\hat{I}^{\dag }(t)$ are required to serve as an orthonormal basis. In the
considered model with unbroken $PT$-symmetry\cite{Benderajp}, the invariant
operators possess real eigenvalues and the average energies of Hamiltonian
are also real. We obtain the LR phase and non-adiabatic Berry phase, which
reduces to the adiabatic one in the slowly varying limit. The non-unitary
time-evolution operator is formulated explicitly. The ket ( $\left\vert \psi
\left( t\right) \right\rangle $) and bra ($\left\langle \psi \left( t\right)
\right\vert $) states evolve with time respectively by the evolution
operators $\hat{U}\left( t,0\right) $ and $\hat{U}^{\dag }\left( t,0\right) $%
. The inner product of two states are evaluated with the help of a metric
operator $\hat{\chi}\left( t\right) $. This model Hamiltonian can be
realized by periodically driven harmonic oscillator.

\section*{Acknowledgements}

This work was supported by the National Natural Science Foundation of China
(Grant Nos. 11275118 and 11874246) and the Science and Technological
Innovation Programs of Higher Education institutions in Shanxi Province
(STIP) (Grant No. 2020L0586).

\end{document}